\documentclass{JHEP3}

\keywords{Classical Theories of Gravity, Brane models}
\preprint{}

\usepackage{graphicx}
\usepackage{latexsym}
\usepackage{amsmath}
\usepackage{amsfonts}
\usepackage{amssymb}

\newcommand{\be}{\begin{equation}}
\newcommand{\ee}{\end{equation}}
\newcommand{\ben}{\begin{eqnarray}}
\newcommand{\een}{\end{eqnarray}}

\newcommand{\bb}{\bibitem}

\title{Gravity on the Bloch Brane}

\author{Adalto R. Gomes\\ Departamento de Ci\^encias Exatas,
Centro Federal de Educa\c c\~ao Tecnol\'ogica do Maranh\~ao,
65025-001 S\~ao Lu\'\i s, Maranh\~ao, Brazil\\ E-mail:
{argomes@cefet-ma.br}}

\abstract{Bloch branes were introduced previously and are
constructed in a system described by two real scalar fields
coupled with gravity in (4, 1) dimensions in warped spacetime
involving one extra dimension. This work investigates gravity on such thick
branes with internal structure and focuses on the effects of massive
graviton modes localization on the brane and to what extent they
might reproduce the 4d gravity at a scale before escaping into the
extra dimension. In
this way gravitational measurements on the brane could reveal the
existence of the extra dimension on some scales, with possible
applications on brane cosmology.
 \\

\vspace{1cm}

}

\begin{document}

\section{Introduction}

Domain walls as extended defects in field theory have been used in
high energy physics for representing brane scenarios with extra
dimensions  \cite{rub,vis,rs2}. Considered as domain walls with an
internal structure, the Bloch brane \cite{bg} is constructed in a
$(4,1)$ model of two scalar fields coupled with gravity. The
interaction among the fields depends on a real parameter, which
generalizes the standard $\phi^4$ model, changing the way the
scalar field self-interacts. In this way the brane is generated
dynamically and the asymptotic bulk metric is a slice of an
$AdS_5$ space. This and other thick brane \cite{gremm,melf} models
are alternatives to the infinitely thin brane models (see, for
instance, Ref. \cite{rs2} and extensions),
 there constructed with non-dynamical source terms in the action.
 
In a previous work \cite{bg} the internal structure of the Bloch
brane was analyzed in terms of the matter energy density as a
function of the extra dimension. It was found that in a certain
region of the coupling parameter, an splitting of the defect in
two interfaces occurs. In that work the effect of gravity on the
internal structure of the domain wall was made evident as an
induced attraction between the internal interfaces that form the
defect.

This intriguing possibility has led us to refine the former investigations and search for resonance
 states \cite {csaki2} for our $r$ small. We found that this occurs in the region of parameter where
 the brane splitting is more accentuated as we report on section 3. In the same section we considered
 the Newtonian potential for the whole range of the $r$ parameter with the aim to study gravity localization.
  This is the main issue of the present work. Before studying this problem, however, let us first
  review some of the results of the Bloch brane model. It is described by the action
\ben
\label{action}
S=\int d^4x\,dy\sqrt{|g|}\Bigl[-\frac14 R+
\frac12\partial_a\phi\partial^a\phi+\frac12\partial_a\chi\partial^a\chi-
V(\phi,\chi)\Bigr]
\een
where $g=\det(g_{ab})$ and the metric
\ben
ds^2&=&g_{ab}dx^adx^b\nonumber
\\
&=&e^{2A}\eta_{\mu\nu}dx^{\mu}dx^{\nu}-dy^2
\een
describes a background with 4-dimensional Poincare symmetry with $y$ as the extra dimension.
 Here $a,b=0,1,2,3,4,$ and $e^{2A}$ is the warp factor. We suppose that
the scalar fields and the warp factor only depend on the extra coordinate $y$.

The action given by Eq.(\ref{action}) leads to the following coupled differential equations for
 the scalar fields $\phi(y)$, $\chi(y)$ and the function $A(y)$ from the warp factor:
\ben
\phi^{\prime\prime}+4A^\prime\phi^\prime&=& \frac{\partial
V(\phi,\chi)}{\partial\phi}
\\
\chi^{\prime\prime}+4A^\prime\chi^\prime&=& \frac{\partial
V(\phi,\chi)}{\partial\chi}
\\
A^{\prime\prime}&=&-\frac23\,\left(\phi^{\prime2}+\chi^{\prime2}\right)
\\
A^{\prime2}&=&\frac16\left(\phi^{\prime2} +
\chi^{\prime2}\right)-\frac13 V(\phi,\chi)
\een
where prime stands for derivative with respect to $y$.

The potential used is
\ben\label{gpot}
V(\phi,\chi)=\frac18 \left[\left(\frac{\partial
W}{\partial\phi}\right)^2 + \left(\frac{\partial
W}{\partial\chi}\right)^2 \right]-\frac13 W^2
\een
where
\ben\label{w}
W(\phi,\chi)=2\phi-\frac23\phi^3-2r\phi\chi^2,
\een
and $r$ is a real parameter.
The particular relation from Eq.(\ref{gpot}) between $V$ and $W$ leads to a description of the
 scalar fields in terms of a coupled set of first-order differential equations, with analytical
 expressions for $\phi(y)$ and $\chi(y)$ identical to the obtained in Refs. \cite {bs,bb} for flat spacetime.

The first-order differential equations which also solve the equations of motion are \cite{df,st}
\ben\label{e1a}
\phi^{\prime}=\frac12\,\frac{\partial W}{\partial\phi}
\een
\ben
\chi^{\prime}=\frac12\,\frac{\partial W}{\partial\chi}
\een
and
\ben\label{Aprime}
A^\prime=-\frac13\,W.
\een
Similar investigations where also done in Ref. \cite{cvetic}.

In the present case the former equations can be solved
analytically leading to the following results \ben\label{phi}
\phi(y)=\tanh(2ry), \een \ben\label{chi}
\chi(y)=\sqrt{\left(\frac1{r}-2\right)}\;{\rm sech}(2ry) \een and
\ben\label{warp}
A(y)\!\!&=&\!\!\frac1{9r}\Bigl[(1-3r)\tanh^2(2ry)-2\ln\cosh(2ry)\Bigr]
\een Graphical analysis \cite{bg} from these solutions for several
values of the parameter $0<r<0.5$ show that the thickness of the
defect increases with decreasing $r$, with $A(0)=0$ implying
canonical normalization for the 4-dimensional metric in the
transverse space on $y=0$ \cite{csaki}. Note also from Eq.
(\ref{chi}) that for the limit $r=0.5$ the one-field scenario is
recovered.

\section{Fluctuations and massive modes}

We consider now the effective 4-dimensional gravitational fluctuations in the
 conformally flat background discussed previously, as well as that of scalar fields around
  solutions (Eqs.\ref{phi}-\ref{warp}):
\ben ds^2=e^{2A(y)}(\eta_{\mu\nu}+\epsilon h_{\mu\nu})dx^\mu
dx^\nu-dy^2 \een and we set $\phi\rightarrow\phi
+\epsilon\tilde{\phi}$ and $\chi\rightarrow\chi
+\epsilon\tilde{\chi}$, where $h_{\mu\nu}=h_{\mu\nu}(x,y)$,
$\tilde\phi=\tilde\phi(x,y)$ and $\tilde\chi=\tilde\chi(x,y)$
represent small perturbations. On the transverse traceless gauge
the metric perturbation separates from the scalars \cite{df},
leading to \ben\label{h} {\bar
h}_{\mu\nu}^{\prime\prime}+4\,A^{\prime} \,{\bar
h}_{\mu\nu}^{\prime}=e^{-2A}\,\Box\,{\bar h}_{\mu\nu}, \een with
$\Box$ the $(3,1)$-dimensional d'Alembertian.

This equation can be decoupled by separating the 4-dimensional plane wave
perturbations in a form of plane waves from the extra dimension contribution.
 We introduce a new variable $z$ that turns the metric into a conformal one. This changes
 the equation for the extra dimension contribution of the metric perturbations in
  a Schroedinger-like form, where no single derivative terms are present. The new
  conformal coordinate $z$ is defined by $dz=e^{-A(y)}dy$. The separation of variables taken as
\ben
{\bar h}_{\mu\nu}(x,z)=e^{ip\cdot
x}e^{-\frac{3}{2}A(z)}\psi_{\mu\nu}(z),
\een
turns Eq.(\ref{h}) into a Klein-Gordon equation for the 4-dimensional
 components of the transverse-traceless ${\bar
h}_{\mu\nu}$, with the remaining Schroedinger-like equation
\ben\label{se}
-\frac{d^2\psi_m(z)}{dz^2}+V_{sch}(z)\,\psi_m(z)=m^2\,\psi_m(z) \een
for its fifth-dimensional component, where we dropped the $\mu\nu$ indices for the wavefunctions,
now labeled by their corresponding energy $m^2$. Here the potential is given by
\ben
\label{Uz}
V_{sch}(z)=\frac32\,A^{\prime\prime}(z)+\frac94\,A^{\prime2}(z)
\een
Particularly interesting is that in this new conformal variable the appropriate inner-product for
 the wavefunctions $\psi(z)$
is the conventional quantum mechanical one \cite{csaki}; in this way we can interpret a
 normalized $|\psi_m(z)|^2$ as the probability for finding a massive mode at the position labeled by $z$.

The solutions of Eq.(\ref{se}) for general $m$ form a tower of states associated with the massive modes
 responsible for the gravitational interaction in the semiclassical theory we are considering.
 Note that the l.h.s. of Eq.(\ref{se}) can be written as a positive definite Hermitean operator, meaning
  absence of tachyonic excitations \cite{bfg} with a zero mode (normalized $(m=0)$ state) given
  by \ben \label{modozero}\psi_{0}(z)=Ne^{3A(z)/4}, \een
where $N$ is a normalization factor.

Graphical analysis from Eq.(\ref{Uz}) shows that for some region of parameters $V_{sch}(z)$
is a ``volcano-type potential" with a
true minimum at $y=0$ for $r^*<r<0.5$, whereas for $r<r^*$ there appears a splitting of the potential
with two separate minima. In Fig. \ref{figUz} we reproduced the figure from Ref. \cite{bg} in order
 to ease the comparison with the results from next section. As the potential goes to zero as
 $|z|\rightarrow\pm\infty$, the possibility for the model to reproduce 4-dimensional gravity can
  be investigated. We are particularly interested on massive gravity on the Bloch brane, mainly on
   the effect of the massive spectrum of excitations for the localization of gravity on the brane.
    The form of the potential shows that besides the zero mode there are no other normalized states,
     with the massive states escaping from the brane as plane waves.

\FIGURE{\includegraphics[{angle=270,width=9cm}]{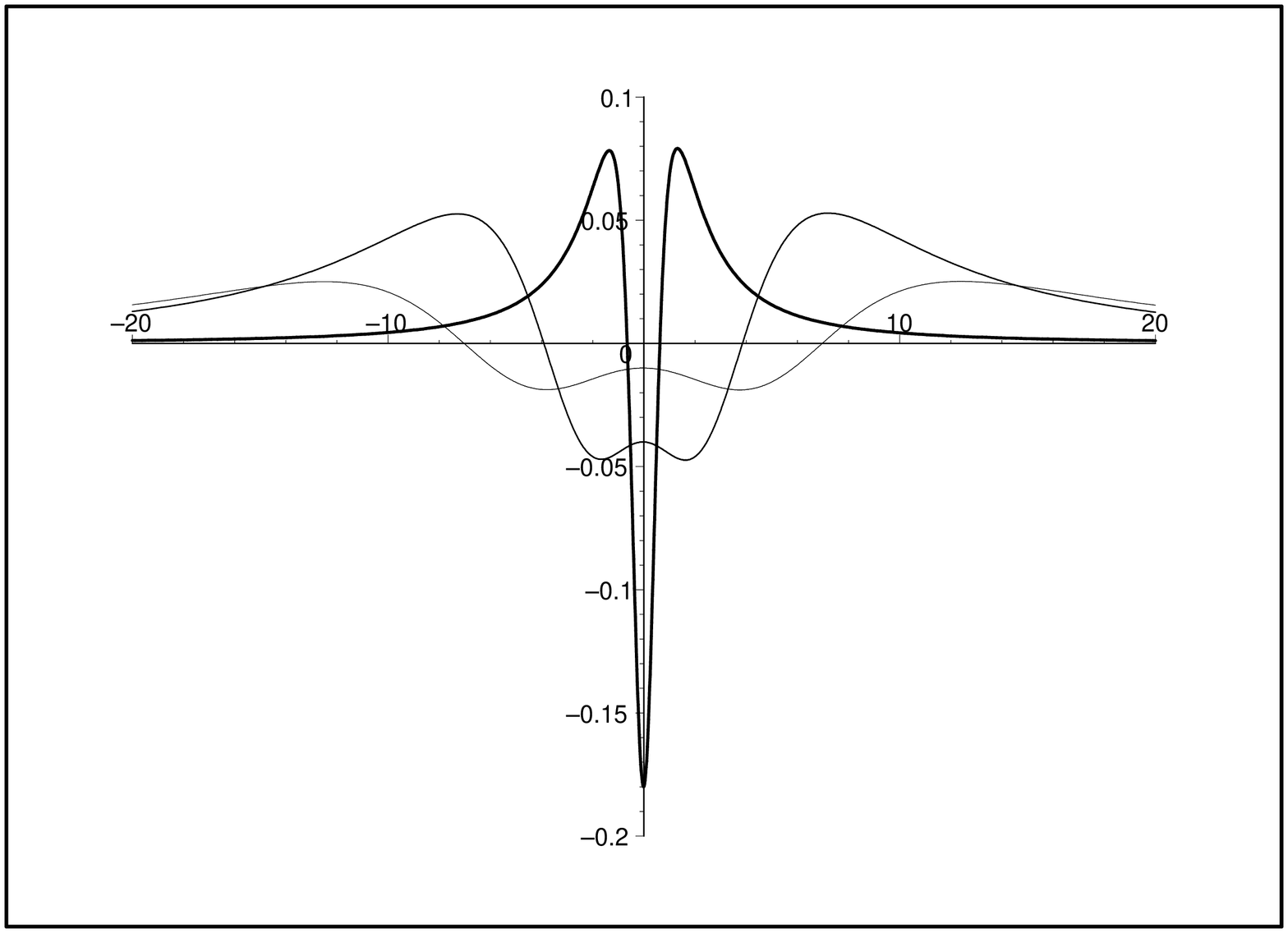}
\caption{Plots of the Schroedinger-like potential $V_{sch}(z)$ for $r=0.05$ (thinner trace) and
 $r=0.10$. Case $r=0.30$ (thicker trace) corresponds to 1/2 $V_{sch}(z)$ to ease comparison.}
\label{figUz}
}

The zero mode is responsible for gravity localization on the
brane, whereas the massive modes
 introduce modifications on the Newtonian potential where in some scale gravity
  on the brane turns out to be essentially 5-dimensional.

We numerically studied the Newtonian potential for two unit test masses separated by a distance $R$.
In order to obtain the
Newtonian potential one sums the tower
of Kaluza-Klein excitations to the usual contribution coming from the zero-mode as \cite{csaki} \ben
\label{UR}
U(R)=G \frac1{R} + \frac 1{M_*^3}\int_0^{\infty}{dm \frac{e^{-mR}}R |\psi_m(0)|^2},
\een where the 4-dimensional coupling $G=M_4^{-2}$, $M_*$ is the fundamental 5-dimensional
 Planck scale and the integration is the considered at the brane position $z=0$ on the thin brane case.
  Separating from the Einstein-Hilbert action the 4-dimensional part from the extra dimensional one leads
   to an expression relating the two scales as \cite{csaki}
\ben
M_4^2=M_*^3\int_{-\infty}^{+\infty} dz e^{-3A(z)/2}.
\label{eqMM}
\een
In this way one can see that those scales are related to the integral of the squared wavefunction for the
 zero-mode - as we have imposed $\psi_0(0)=1$ we cannot associate a true probability for $\psi_0(z)$. That
  is the reason one can say integrability of the zero-mode is equivalent to a non-null 4-dimensional
  gravitational constant.
Note that this does not guarantee localization of gravity in all scales, as the massive modes can modify
 sensibly the gravitational potential on the brane. However, see from Eq.(\ref{UR}) that the higher massive
  modes are exponentially suppressed due to the Yukawa coupling. In this way gravity localization
  and the reproducibility of 4-dimensional gravity on the brane is strongly related to the
  decoupling of the soft KK modes on the brane. In order to make easier the numerical approach to
   obtain the function $U(R)$, see that the integral on the continuum modes is already saturated at $m\sim 10/R$.

Eq.(\ref{eqMM}) shows that $|\psi_m(0)|^2$ is of fundamental
importance in order to achieve a true description of the Newtonian
potential. As the massive modes are on the form of plane waves,
normalization can be considered in a box of fixed length. The
conformal transformation from $y$ to $z$ is done in order to
achieve a distribution
 of points on the $z$ variable with fixed step with a controlled precision. With the function $y(z)$
  one can obtain $A(z)$ and the Schroedinger potential $V_{sch}(z)$. The massive modes are obtained using
   the Numerov method \cite{mh}, whose efficiency was evaluated after comparison of the $m=0$ case with
   the expression given by Eq. (\ref{modozero}) for the zero-mode. For each value of the parameter $r$ we
    have chosen a box extending from $z=-z_{max}$ to $z=z_{max}$, with $z_{max}$ chosen for each $r$ such that
    the potential $V_{sch}$ was near to the asymptotic value zero. This and a sufficiently large number
     of points turns out to be necessary in order to obtain better efficiency for comparison with the
      zero mode. We also expect that the normalization procedure for higher massive modes turns out to
       have higher confidence as the box collects larger number of wavelengths.
We conclude from Fig.  \ref{figUz} that for lower values of $r$ one needs boxes with larger $z_{max}$.
The Numerov method consists of a discretization of the Schroedinger equation in order to get a recurrence
 formula after choosing a fixed value for the non-normalized wavefunction at $z=0$.
{\small
\begin{center}
        \begin{tabular}
        {|l |c|r|}\hline r & $M_4^2/M_*^3$ & $|\psi_0(0)|^2 $\\
        & & (Eq. \ref{modozero})\\
        \hline 0.05 & 14.67 &  0.0681 \\
        \hline 0.10 & 8.85 &   0.113 \\
        \hline 0.15 & 6.62 & 0.151 \\
        \hline 0.20 & 5.43 & 0.184 \\
        \hline 0.30 & 4.13 & 0.241 \\
        \hline 0.50 & 3.04 & 0.329 \\
        \hline
        \end{tabular}
   \end{center}}
{\small {\bf Table I: }{Ratio between 4-dimensional coupling and
fundamental 5-dimensional Planck scale for some values of r. Note
the greater importance for $M_4$ for smaller values of $r$. Third
column contains the normalized zero-mode on the brane obtained
using Eq. \ref{modozero}. This can be compared with the
corresponding values for $m=0$ obtained with Numerov method and
displayed on Fig. \protect\ref{psisq}}.} \\

Fig. \ref{psisq}  shows the normalized probability for massive modes on the brane as a function
 of the masses of the modes for $r=0.3$ and $r=0.05$. First note that in the limit $m\to 0$, for
 the chosen parameters displayed on the same figure, Numerov method gives $|\psi_0(0)|^2=0.247$
 for $r=0.3$ and $|\psi_0(0)|^2=0.0690$ for $r=0.05$. Comparing those values with Table I, this
  results in a relative error for determining the zero-mode of $2.5\%$ and $1.3\%$, respectively.
  This shows that our method has higher precision for lower values of $r$. In particular, for higher
   values of $r$ this error is difficult to reduce considerably as the Schroedinger potential goes
   slowly to zero in comparison to the volcano-like potential for lower values of $r$.
Note also that the near-zero mode scattering states occur on the brane at higher probability,
with the higher massive modes tending to a fixed probability. The asymptotic behaviour for higher
 massive modes shows that exponential suppression on Eq.(\ref{UR}) is effective. The appearance of
  resonances could be identified as isolated peaks on energies below the maximum peak of the volcano
   potential and be responsible for a modification of the Newtonian potential obtained further in
    this section.  From the figure one can see few evidence of resonances for $r=0.3$. However, for
    $r=0.05$ there is an oscillation on the curve for small masses, indicating the presence of resonances
     for values of $r$ smaller than a critical one. In this way we see that the splitting of the Schroedinger
      potential for lower values of $r$, besides leading to an internal structure \cite{bg} on the brane,
      also leads to a richer distribution of the massive modes. Extensions of Fig. \ref{psisq} for higher
      masses up to $m=1000$ show that the probabilities tend to constant values on both cases, indicating
       similar contributions to the 4-dimensional gravity on short distances, as we will see in the following.

\FIGURE{\includegraphics[{angle=0,width=7.5cm}]{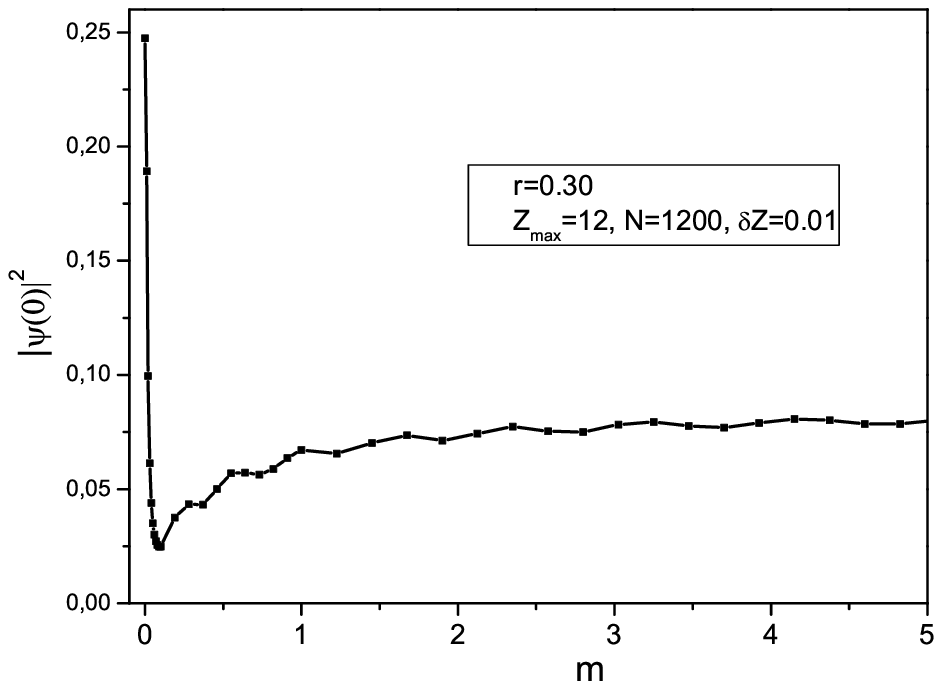}
\includegraphics[{angle=0,width=7.5cm}]{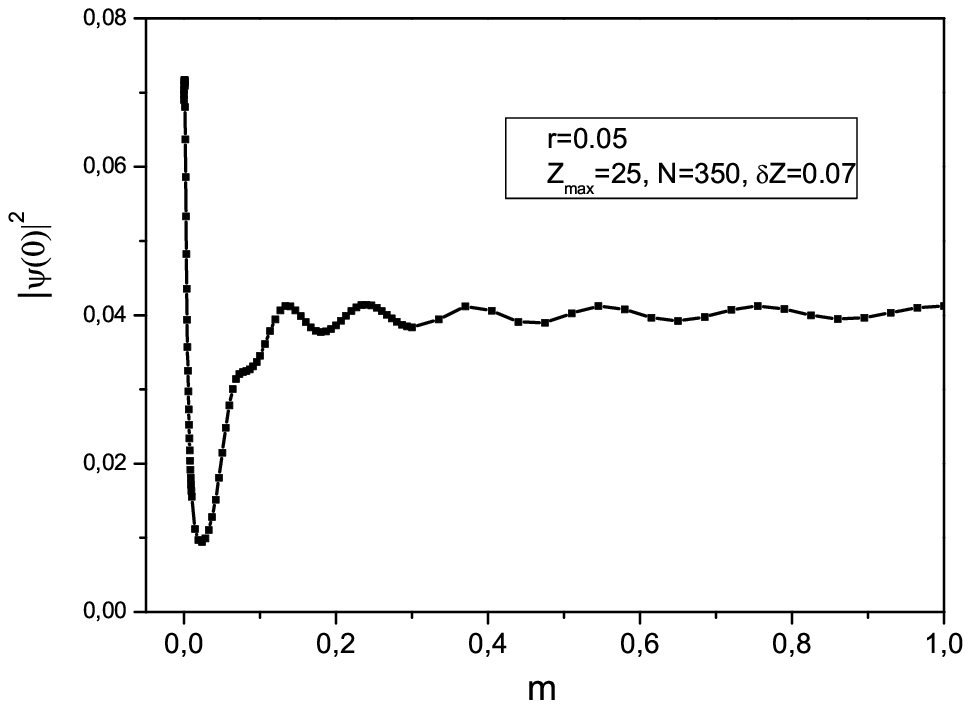}
\caption{Normalized probability for finding massive modes on the brane as a function of the mass of
the modes, for $r=0.3$ (left) and $r=0.05$ (right). $Z_{max}$ is for box thickness, $N$ the number of
 points and $\delta z$ the step  on $z$ variable.}
\label{psisq}
}

To better quantify the results, we are
interested on the $L$ parameter, defined as the exponent of the
R-dependence of the Newtonian potential,
\ben
\label{URL}
U(R)=\frac A {R^L}, \een where $A$ is independent of $R$, and is
related to the 5-dimensional gravitational constant $G_5$. From the former equation we get
 $L=-\partial (\log U)/\partial(\log R)$.
The dependence of the exponent $L$ as a function of $\log R$ has an interesting interpretation.  It describes
the dimensionality of the gravitational interaction as a function of the scale of separation between two test
 masses.

Note, however, from Eq.(\ref{UR}) that in order to obtain the Newtonian potential, an important information is
 the relation between the two scales $M_4^2$ and $M_*^3$. Table I displays some results for their ratio depending
  on the $r$ parameter. Note from the Table that the 4-dimensional coupling constant $G=1/M_4^2$ is more pronounced
  for larger values of $r$, indicating that the gravity localization is favored for those parameters. In particular,
   we see that when the field $\chi$ is tuned to zero for the limit $r=0.5$ we see that $G$ achieves the maximum
    value compared to the fundamental scale. In this way we can associate the introduction of the additional
    field $\chi$ and the consequently richer structure of the brane to a larger modification of 4-dimensional gravity
     due to the extra dimension.

Some results for $L$ as a function of $\log R$ can be seen in Fig. \ref{seqLlogR} for coupling parameter $r=0.05$
 and $r=0.3$. From the figure one can see that the exponent $L$ interpolates between the usual 4-dimensional
  gravity $L=1$ to a pure 5-dimensional gravity $L=2$. The extra-dimensional characteristic is revealed on the
   brane for short separation between the test masses, whereas the usual Newtonian potential is recovered for the
   whole scale of distances above a certain transition region. In Fig. \ref{seqLlogR} one can see that the explicit
    contribution for $L$ coming from the continuum modes would induce a transition region with higher and lower values
     around the $L=2$ region. In this way one can see that in the present model the normalized zero mode is essential
      for a proper recovering of the 4-dimensional gravity on the brane. Note also that on the transition region,
       one gets a higher dimensionality for lower values of $r$. If we compare this with Fig. \ref{figUz} we see
        that the splitting of the Schroedinger potential favors the extra dimension to be revealed on the brane.

Note that the Bloch brane scenario leads to different limit than RS \cite{rs2} model for the Newtonian
 potential on short distance scales. As one knows, the $AdS$ case for RS model gives $L=3$ for those scales.
  However, the continuum modes decouple and the zero-mode is normalized in both scenarios, leading to corrections
   to Newton's law that are suppressed on large distance scales.

\FIGURE{\includegraphics[{angle=0,width=11cm}]{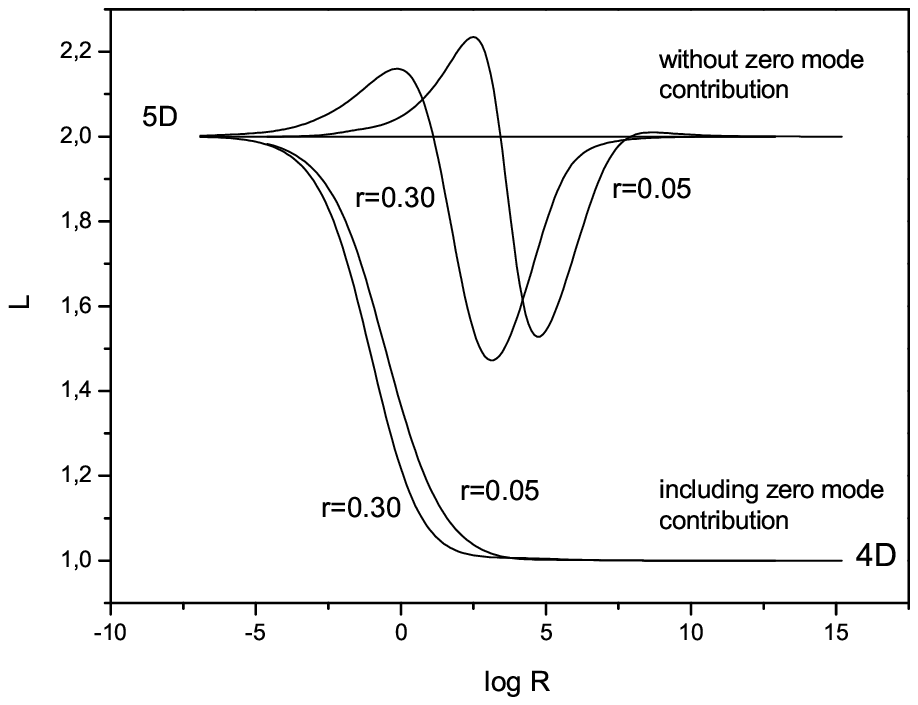}
\caption{Gravity localization on the brane for $r=0.3$ and $r=0.05$: for large scales the Newtonian potential
 is essentially 4-dimensional.  }
\label{seqLlogR}
}

\section{Conclusions}
In this work we have studied the Bloch brane, with particular interest on signals from the extra dimension
 on the Newtonian potential. The spectrum analysis for the massive modes indicates a richer structure for low
  values for the parameter $r$, where some resonances do appear. This richer structure can be related to the
   splitting of the Schroedinger potential observed on Ref. \cite{bg}. Also the resonance structure was revealed
    due to an improvement of the numerical  normalization of the massive modes in a box and increased
    the understanding of the Bloch brane. Our results for the ratio $M_4^2/M_*^3$ indicate that modifications
     coming from the extra dimension turns to be more important for lower values of the $r$ parameter.
      This observation was confirmed from the Newtonian potential analysis where we have considered
      the change of the potential decay law $U\sim 1/R^L$ with the increase of the separation $R$ between
       two unit masses. We analyzed numerically the continuous change of the exponent $L$ with $\log(R)$
        indicating the change of a pure 4-dimensional character for the gravitational interaction on the
         brane to a 5-dimensional one as one moves to shorter separation between the masses. In this way
          our result addresses the effort for experimental determination of gravitational interaction on
          short distances. In the Bloch brane scenario experimental modifications for the usual Newtonian
           law would indicate the presence of the extra dimension, for the parameter $r$ very small.

\section{Acknowledgements}
The author thanks D. Bazeia and F.A. Brito for suggestions and
stimulating discussions and CNPq for financial support.


\begin{thebibliography}{99}

\bb{rub}V.A. Rubakov and M.E. Shaposhnikov, Phys. Lett. B 125, 136 (1983).
\bb{vis}M. Visser, Phys.Lett. B 159, 22(1985).
\bibitem{rs2}L. Randall and R. Sundrum, Phys. Rev. Lett. {\bf83},
4690 (1999); [arXiv:hep-th/9906064].
\bibitem{bg}  D. Bazeia and {{ A.R. Gomes}}, JHEP {\bf0405} (2004) 012;
[hep-th/0403141].
\bb{gremm}M. Gremm, Phys. Lett. B 478, 434 (2000); hep-th/9912060.
\bb{melf}A. Melfo, N. Pantoja, and A. Skirzewski, Phys. Rev. D 67, 105003 (2003); gr-qc/0211081.
\bibitem{csaki2}C. Csaki, J. Erlich, and T. Hollowood, Phys.Rev.Lett. 84 (2000) 5932-5935;
[arXiv:hep-th/0002161].
\bb{bs}D. Bazeia, M.J. dos Santos and R.F. Ribeiro, Phys. Lett.
A{\bf208}, 84 (1995).
\bb{bb}D. Bazeia and F.A. Brito, Phys. Rev. D {\bf61}, 105019 (2000).
\bb{df}O. DeWolfe, D.Z. Freedman, S.S. Gubser, and A. Karch,
Phys. Rev. D {\bf62}, 046008 (2000).
\bb{st}K. Skenderis and P. Townsend, Phys. Lett. B {\bf 468}, 46 (1999); [arXiv:hep-th/9909070].
\bb{cvetic}M. Cvetic and H.H. Soleng, Phys. Rept. 282 (1997) 159; [arXiv:hep-th/9604090].
\bibitem{csaki}C. Csaki, J. Erlich, T. Hollowood and Y. Shirman,
Nucl.Phys. B581 (2000) 309-338; [arXiv:hep-th/0001033].
\bb{bfg}D. Bazeia, C. Furtado and A.R. Gomes, J. Cosmol. Astropart. Phys. {\bf 0402} (2004) 002; [Arxiv: hep-th/0308034].
\bb {mh} M. Mueller and H. Huber, {\it Solution of the 1-D Schroedinger Equation}, http://www.mapleapps.com


\end{thebibliography}
\end{document}